\documentclass[runningheads]{llncs}
\usepackage{graphicx}
\usepackage[caption=false]{subfig}
\begin{document}
\title{DeepQuarantine for Suspicious Mail}
\author{ Nikita Benkovich \and
Roman Dedenok \and
Dmitry Golubev}
\authorrunning{N. Benkovich et al.}
\institute{Kaspersky, Moscow 125212, Russia \\
\email{\{Nikita.Benkovich, Roman.Dedenok, Dmitry.S.Golubev\}kaspersky.com}}
\maketitle              % typeset the header of the contribution
\begin{abstract}
In this paper, we introduce DeepQuarantine (DQ), a cloud technology to detect and quarantine potential spam messages. Spam attacks are becoming more diverse and can potentially be harmful to email users. Despite the high quality and performance of spam filtering systems, detection of a spam campaign  can take some time. Unfortunately, in this case some unwanted messages get delivered to users. To solve this problem, we created DQ, which detects potential spam and keeps it in a special Quarantine folder for a while. The time gained allows us to double-check the messages to improve the reliability of the anti-spam solution. Due to high precision of the technology, most of the quarantined mail is spam, which allows clients to use email without delay. Our solution is based on applying Convolutional Neural Networks on MIME headers to extract deep features from large-scale historical data. We evaluated the proposed method on real-world data and showed that DQ enhances the quality of spam detection.

\end{abstract}
\section{Introduction}
Nowadays it is hard to imagine a life without e-mail communication, particularly in business area. The growth of e-mail's popularity is accounted for low cost and high effectiveness of exchanging messages. The same factors contribute to the increasing amount of spam.
 According to a report by Kaspersky~\cite{ref_securelist}, the average percentage of spam in the global mail traffic in Q1-Q2 2019 was 57.64\%, up 1.67 p.p. compared to the previous reporting period. The largest share of spam was recorded in May ($58.71\%$). In Q2 2019, Kaspersky alone detected more than 43 million of malicious email attachments and about 130 million phishing attacks. Statistics show that spam campaigns are a serious threat these days.
 A large amount of spam in the mailbox causes a decrease in performance, wastage of storage space and inconvenience when using e-mail. Moreover, spam messages can carry malicious content, phishing and fraud schemes, which can harm both casual users and business around the world.

Anti-spam software companies aim to protect users against malicious mail, and, crucially, ensure the delivery of all legitimate messages to them.
Otherwise, even one misclassified message, for example, from a business conversation, can lead to significant reputation risks. To reach a low false positive rate, anti-spam decisions must be very reliable, which obviously reduces detection rate. To solve this problem, commercial anti-spam products delay potential spam messages to recheck them after a certain time to improve the reliability of the anti-spam solution. The Axway Inc. in~\cite{ref_patent} described delay technique in e-mail filtering system, which provides a store and the transmission path of quarantined data. This mechanism was established reliable and now its different modifications are used in many companies such as Cisco, Barracuda and others. 
 
In this paper, we describe a novel approach to quarantine messages. Our work focuses on applying Deep Learning~\cite{ref_dl} techniques on MIME (Multipurpose Internet Mail Extensions)~\cite{ref_mime}  headers to classify potential spam. Unlike most research papers, our solution does not process body content of a message. The proposed architecture has three inputs: a char sequence of Message-Id, a sequence of headers and X-Mailer. For extracting information from sequential data, we use one-dimensional convolutional neural network (CNN). This method was applied on characters to text classification~\cite{ref_clcn}. It has been shown that this approach can be competitive to traditional solutions for example with a simple long-short term memory net (LSTM)~\cite{ref_lstm}. Moreover, CNNs do not depend on the computations of the previous states unlike LSTM. This fact affects model performance, which is extremely important in real-time services.

We evaluated our approach on a large-scale dataset. In the experiments, we showed that combination of our proposed model and traditional spam filters improves in classification rates.

\section{Related Work}
Cybercriminals continue to look for new ways to spread spam and improve previous techniques. Traditional signature approaches are becoming less effective compared to previous years. The reasons are poor generalization ability and the need to use human resources to find new attacks and develop signatures to block them.

Machine learning techniques have recently become very effective to fight spam.
Most research papers propose different methods to handle body content of a message.~\cite{ref_pois_attacks} suggested a defence strategy against poisoning attacks, when spammers enrich messages with legitimate words to defeat filters. They showed that bagging ensembles could be very promising in this task.
In~\cite{ref_Dhamani} authors applied deep learning and transfer learning techniques to detect different attacks such as phishing, social engineering, propaganda and others.~\cite{ref_pois_attacks} demonstrated a phishing content classifier based on a recurrent neural network.

There are also related works that use non-content features for spam detection.~\cite{ref_Zhang} noted that message headers are a powerful source of features for spam filtering. The experiments showed that using only features from headers could achieve comparable or better performance than where using body content. In~\cite{ref_Qaroush} and~\cite{ref_Hu} authors proposed hand-crafted methods to extract features from e-mail headers, and evaluated performance of various machine learning classifiers using a prepared corpus.

Publicly available benchmark datasets on e-mail spam highlighted in~\cite{ref_Bhowmick} are not regularly updated thus do not reflect actual threats. 
Publication of real email collections is almost impossible since this data is associated with numerous confidential and legal restrictions.
Moreover, available datasets can be highly biased because they contain conversations between a small group of users. For example, the popular Enron corpus~\cite{ref_Enron} is deemed to be in the public domain as the result of an investigation after the company's collapse and contains only communications between Enron employees. These factors complicate research in this area and the adaptation of the proposed methods in the real world.

\section{Method}
In this section, we introduce the design of DQ. We describe three main parts of the new technology. First, we focus on backend logic, which is responsible for message transactions and the system-customer relationship. Then we illustrate preprocessing of message headers. Finally, we show design of model for spam classification.

\subsection{Backend logic}

\begin{figure*}[htp]
   \subfloat[Origin-based ]{\label{fig:f1}
      \includegraphics[width=.4\textwidth]{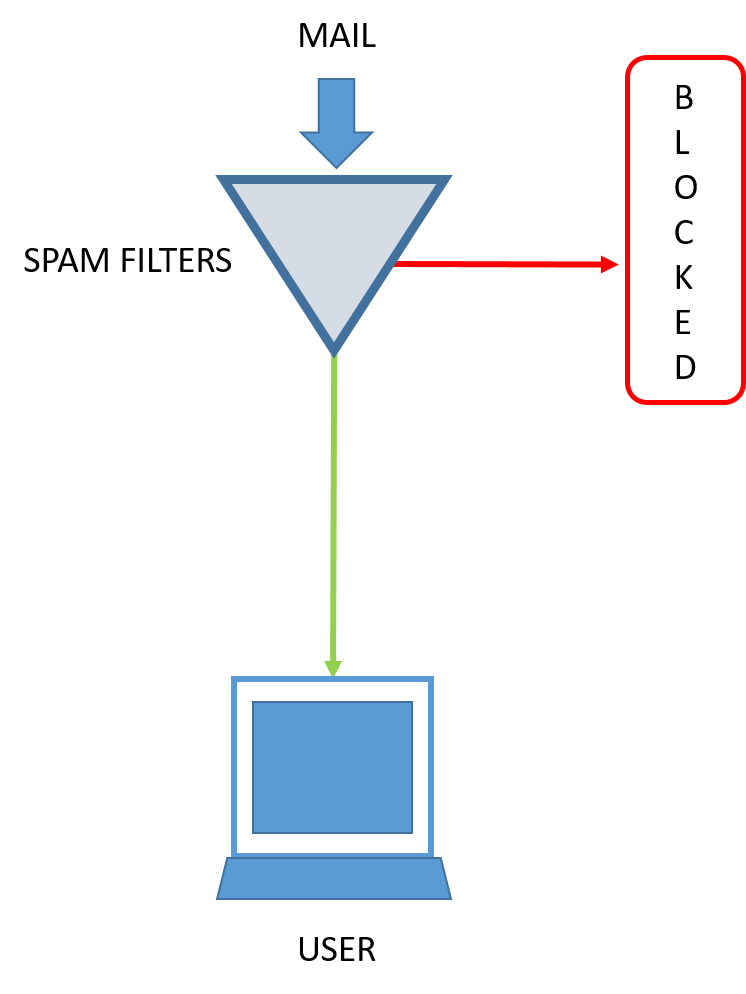}}
~
   \subfloat[DQ implementation]{\label{fig:f2}
      \includegraphics[width=.525\textwidth]{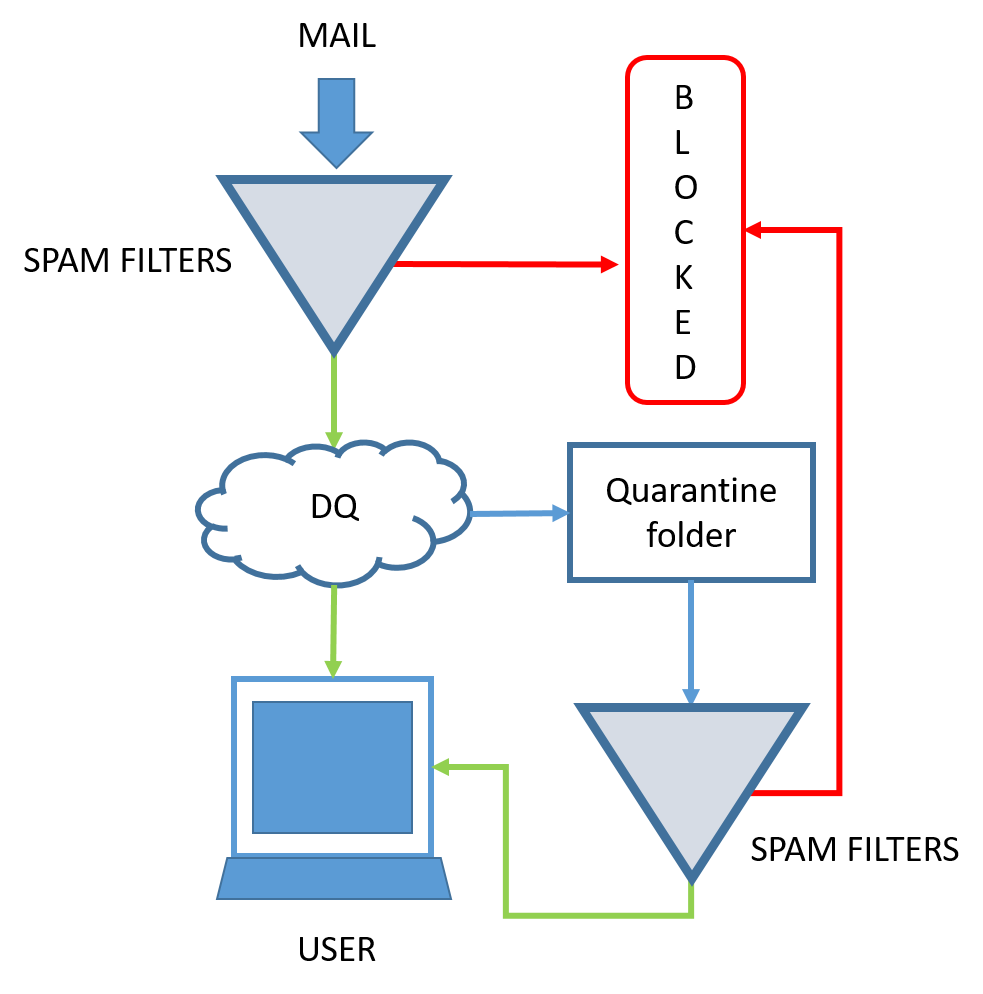}}

   \caption{Backend of spam-filtering systems}
   \centering

\end{figure*}

According to Figure \ref{fig:f1} messages in origin-based scheme are processed by complicated system of spam filters before delivered to user. Moreover, spam filters are regularly updated because statistical properties of spam campaigns change over time. Indeed, this approach can be used and potentially provides high detection rate. In real life, most missed spam is detected shortly after updating filters. Unfortunately, the considered scheme delivers these messages to users because spam decisions are made once when a message is received.

To solve this problem we implemented DQ, illustrated on Figure \ref{fig:f2}.
DQ is a cloud technology, which provides request-response logic with an installed anti-spam service on user's machines. The main objective of DQ is classification of potential spam. 
After a messages passed through filters, the service sends a request to DQ with message headers and waits for a response. Meanwhile, DQ handles input data and returns \textit{true} if message should be delayed or \textit{false} otherwise. 
Of course, in real life organization of this communication to process the big data that accumulates from different user nodes is not a trivial task.
We do not go deep into implementation details and focus on logic of the technology.
As shown on Figure \ref{fig:f2}, suspicious mail is put in the Quarantine folder for a while, others are delivered to user.
When the time is over, quarantined messages pass through filters again. It should be noted that DQ only receives required headers and returns the quarantine decision, all delayed mail in Quarantine folder is located on the user PC.

The proposed scheme allows to gain the time to update filters and double-check suspicious messages to improve the reliability of the anti-spam solution. Moreover, this implementation provides a low-cost way to update the model that is extremely important to adapt to new spam tactics or changing mail transfer protocols.

\subsection{Headers preprocessing}

It is well known that the feature selection plays a big role in model performance. The e-mail provides a large amount of information about a sender and content of the message. Some of this data can be absolutely useless and add unwanted noise that can be a reason of lower model performance. Our solution is based on non-content classification. Due to this fact, we are able to transfer data to cloud service and collect this type of information using simpler way than in content-based case. Another important aspect is the ability to quickly extract features from message. As far as DQ is a real-time service, performance  is very important to ensure email communication without delay. 

At the moment, the model takes: Message-ID, a sequence of message headers (HeaderSeq) and X-mailer. To bypass protection systems and spread malicious mail, spammers often use their  own Mail User Agent (MUA). MUAs are responsible for preparing email messages for transferring to a Mail Transfer Agent (MTA). One of the MUA tasks is to create and fill correct MIME headers. Some of attackers ignore it and can use random content for headers. Others try to fake headers to make them look like real ones. We focus on Message-ID and HeaderSeq for several reasons. Firstly, these features have non-trivial structure. Secondly, the form of Message-Id and the order of headers in HeaderSeq can vary depending on the type of MUA, which creates a tight connection between features. These facts make compromising more difficult, which helps the model to detect spam. We also added X-mailer to define MUA. Below we describe features and their representation for the classifier.

\subsubsection{The Message-ID} provides an identifier for messages and looks like a sequence of US-ASCII characters between an angle bracket pair. For example:
$$\langle \textit{5ced853647da4fd3689a26db412fa4c1@foo.com} \rangle$$
Message-ID consists of two parts splitted by @. The left part of the Message-ID is a hash that has a specific structure for different MUAs. The right part is a domain. 
The Message-ID is transformed to a tensor size $l$, where each row vector is a char embedding. For encoding, we build a vocabulary that maps US-ASCII chars (without special characters) to trainable embeddings. In addition, we added two symbols $<EOS>$ for the end of a string and $<UNK>$ for unknown characters. In case the length of Message-ID is greater than $l$, the first $l$-characters are taken. In case length of Message-ID is less than $l$, the sequence is filled with $<EOS>$ to the length $l$.

\subsubsection{The HeaderSeq} is a sequence of MIME headers in the message. The order of headers can vary depending on the type of MUA. 
The encoding of HeaderSeq has the same scheme as the Message-ID. The only difference is that we operate with header names, not characters. For example:
$$\textit{subject:from:to:date:message-id:content-type:}$$ 
is a possible HeaderSeq. The final representation is a tensor with fixed shape where each row is an encoded header. The number of rows was estimated from statistics as a 95-percentile of HeaderSeq length.

\subsubsection{The X-Mailer} is the name of a MUA. Before encoding, we preprocess the X-Mailer to get information only about the type of MUA. For an actual e-mail program, we drop information about version and release. For example:
$$\textit{Microsoft Windows Live Mail 14.0.8117.416}$$
is transformed to \textit{Microsoft}. This helps significantly reduce the size of the feature space. We also conducted experiments that used the name and version of MUA, but this did not increase performance. For an unknown e-mail program, we created a special category. The encoding is done by using \textit{one-hot encoding}.

\subsection{Classifier design}
In this section, we describe the architecture of the spam classifier. Despite the fact that DQ does not block messages, we cannot delay all of them for re-checking, because this significantly increase the delivery of an e-mail. Moreover, to ensure that e-mail work without delays, DQ has a time limit for the response. If the time is over, the message is delivered to the user without applying DQ. For these reasons, we have a trade-off between model complexity and computation time.

\begin{figure}
\includegraphics[width=\textwidth]{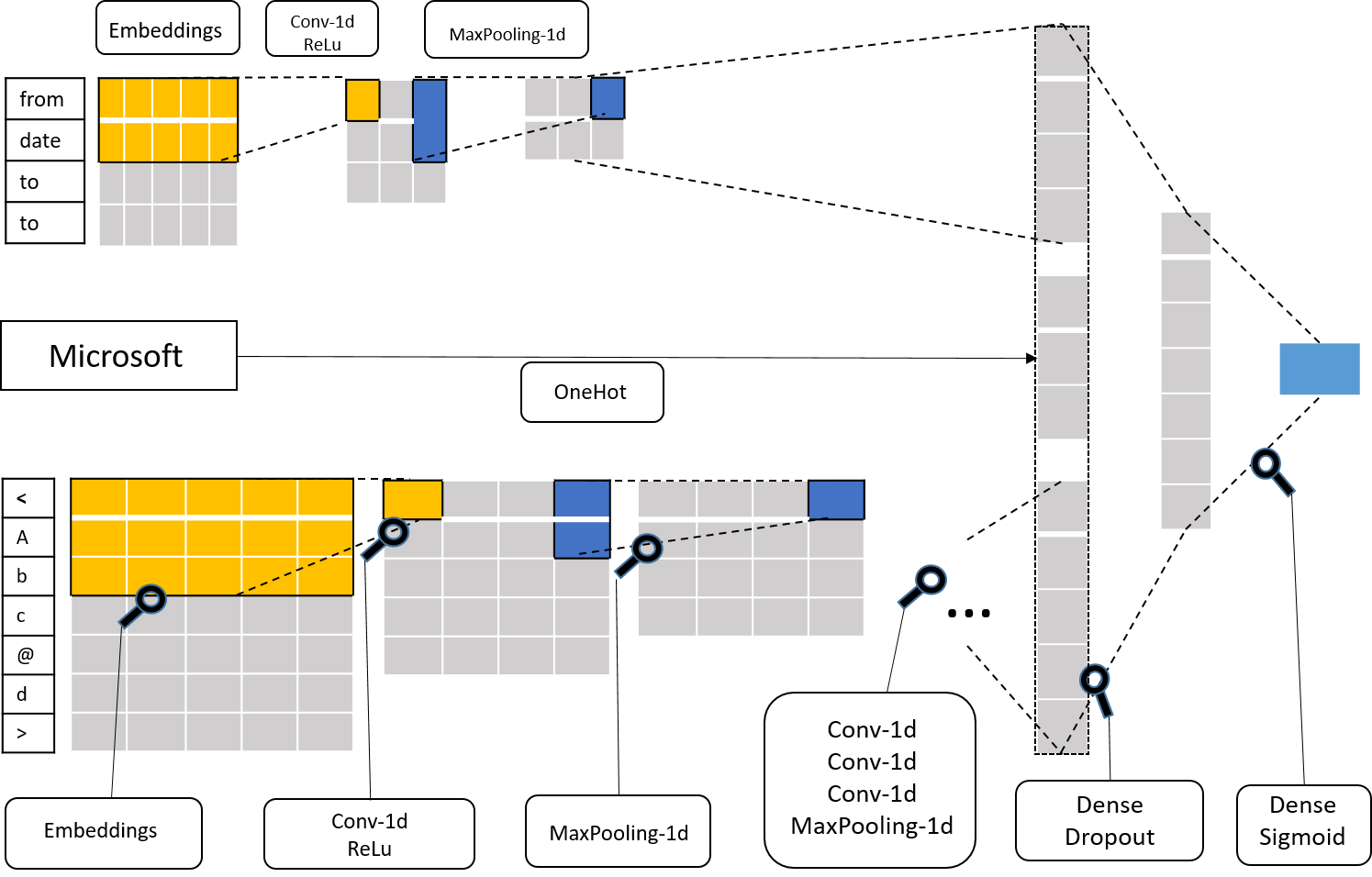}
\caption{Architecture of spam classifier} \label{fig:architecture}
\end{figure}

Figure \ref{fig:architecture} demonstrates the model architecture. Following~\cite{ref_clcn}, for Message-ID and HeaderSeq we applied a temporal CNN to extract features from sequential data. 
This kind of CNN applies convolutional filters along one dimension and capture all units from others. Also we used the one-dimensional version of the max-pooling module applied in~\cite{ref_max_pooling}.

For the Message-ID we designed a subnet with four temporal convolution layers with a fixed number of filters for each of them. We applied relu as activation function. Initially we use a layer with biggest filter size to extract information from longer subsequences. After the first and last layers, we inserted a temporal max-pooling layer to ensure stability of training. 
In the HeaderSeq branch, we used two layers: a temporal convolution layer and a temporal max-pooling layer. The shallow architecture is the result of a small length of HeaderSeq. 
The outputs from the convolutional nets are concatenated with the encoded X-Mailer to a one-dimensional tensor as illustrated in Figure \ref{fig:architecture}. Finally, we added two fully connected layers and inserted a dropout~\cite{ref_dropout} between them for regularization. We used sigmoid activation to obtain the probability of spam for model's output.

\section{Experiments}
In many research papers it is stated that a CNN usually requires large-scale datasets to work and achieve competitve performance in difference areas. Unfortunately, public datasets for spam classification are fairly small and do not show actual threats because they are not regularly updated. 

In this work, we used a collection that consists of metadata from tens of millions of real-time e-mail scans. We split the data into training and test datasets by timestamp to avoid leaking information from the future into the past. We sampled $120$ million objects for training and $40$ million objects for testing. In both datasets, the proportion of spam is about $40$ percent. We optimized weights of the model using SGD with momentum of $0.9$ to minimize the cross-entropy loss. We initialized the model weights using a Gaussian distribution and trained all layers together throughout nine epochs and halve the learning rate every three epochs. 

\begin{figure}
\includegraphics[width=0.6\textwidth]{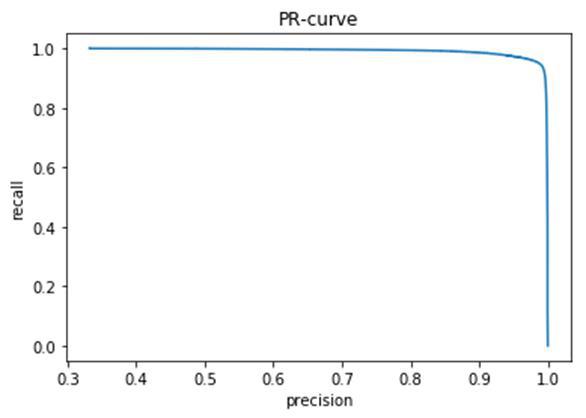}
\centering
\caption{Precision-recall curve computed on the test dataset} \label{fig:pr_curve}
\end{figure}

We show the PR-curve in Figure \ref{fig:pr_curve} to demonstrate the model performance on the test data. As mentioned earlier, a classifier should have high precision to deliver legitimate e-mail messages without delay.
We defined a probability threshold for which the precision is equal $0.998$ and the recall is $0.823$. We tested the DQ with this classifier in the course of 4 weeks in the real world. Our internal tests showed that the proposed technology detects up to $30\%$ of previously missed spam.

\section{Results}

This article proposes a non-content-based classification approach to delay potential spam messages in real time. On the one hand, we demonstrated a novel feature set and way to handle it for a spam classification task.
On the other hand, we show that this method is well-suited for enterprise solutions because it has a simple update scheme, high performance and a low false positive rate. Furthermore, combining this technology with resource-intensive checks that require additional time for verification/response (such as a Whois requests, in-depth content verification, etc), we can get a fast and cost-effective system for detecting spam messages.


\end{document}